\newcommand\beq{\begin{eqnarray}}
\newcommand\eeq{\end{eqnarray}}
\newcommand\ctf{C_{24}}
\newcommand\csf{C_{75}}
\newcommand\cth{C_{200}}
\newcommand\barM{\hat{M}}
\newcommand\barA{\hat{A}}
\newcommand\barm{\hat{m}}
\def\slashchar#1{\setbox0=\hbox{$#1$}           
   \dimen0=\wd0                                 
   \setbox1=\hbox{/} \dimen1=\wd1               
   \ifdim\dimen0>\dimen1                        
      \rlap{\hbox to \dimen0{\hfil/\hfil}}      
      #1                                        
   \else                                        
      \rlap{\hbox to \dimen1{\hfil$#1$\hfil}}   
      /                                         
   \fi}
\newcommand\missingET{{\slashchar{E}_T}}
\def\lsim{\mathrel{\rlap{\lower4pt\hbox{$\sim$}}
    \raise1pt\hbox{$<$}}}                
\def\gsim{\mathrel{\rlap{\lower4pt\hbox{$\sim$}}
    \raise1pt\hbox{$>$}}}

\documentclass[
amsmath,
prd,nofootinbib,floatfix,11pt
]{revtex4}

\allowdisplaybreaks
\usepackage{graphicx}
\usepackage{axodraw}
\usepackage{setspace}

\begin{document}
\renewcommand{\theequation}{\arabic{section}.\arabic{equation}}

\title{\Large%
\baselineskip=21pt
Compressed supersymmetry and natural neutralino dark matter
from top squark-mediated annihilation to top quarks}

\author{Stephen P. Martin}
\affiliation{
Physics Department, Northern Illinois University, DeKalb IL 60115 USA\\
{\rm and}
Fermi National Accelerator Laboratory, PO Box 500, Batavia IL 60510}

\begin{abstract}\normalsize
\baselineskip=15pt
The parameters of the Minimal Supersymmetric Standard Model appear to 
require uncomfortably precise adjustment in order to reconcile the 
electroweak symmetry breaking scale with the lower mass limits on a 
neutral Higgs scalar boson. This problem can be significantly ameliorated 
in models with a running gluino mass parameter that is smaller than the 
wino mass near the scale of unification of gauge couplings. A 
``compressed" superpartner mass spectrum results; compared to models with 
unified gaugino masses, the ratios of the squark and gluino masses to the 
lightest superpartner mass are reduced. I argue that in this scenario the 
annihilation of bino-like neutralino pairs to top-antitop quark pairs 
through top squark exchange can most naturally play the crucial role in 
ensuring that the thermal relic dark matter density is not too large, with 
only a small role played by coannihilations. The lightest superpartner 
mass must then exceed the top quark mass, and the lighter top squark 
cannot decay to a top quark. These conditions have important implications 
for collider searches. \end{abstract}

\maketitle

\tableofcontents

\vfill\eject
\baselineskip=15pt

\section{Introduction}\label{sec:intro}
\setcounter{equation}{0}
\setcounter{footnote}{1}

Softly-broken supersymmetry is a leading candidate to explain the 
hierarchy of the Planck mass scale and other high-energy scales to the 
electroweak symmetry breaking mass scale \cite{hierarchyproblem}. In 
extensions of the Standard Model with a fundamental Higgs scalar, 
obtaining this hierarchy would seem to require tuning of the Higgs squared 
mass parameter to about one part in $10^{32}$. The Minimal Supersymmetric 
Standard Model (MSSM) \cite{primer} solves this problem by introducing 
superpartners with masses near the electroweak scale. In addition, with 
the assumption of $R$-parity conservation, the most dangerous 
(renormalizable) contributions to proton decay are eliminated, and the 
lightest supersymmetric particle (LSP) can serve 
\cite{neutralinodarkmatter}-\cite{DarkSUSY} as the cold dark matter 
required by cosmology \cite{WMAP}-\cite{PDG}.

However, the fact that the CERN LEP $e^+e^-$ collider did not discover a 
Standard Model-like light neutral Higgs scalar boson, 
placing a limit $M_{h^0} > 114$ GeV 
\cite{LEPHiggsbounds}, has put some tension on the allowed parameter space 
in the MSSM. This is because $M_{h^0}$ is bounded above at tree level by 
$m_Z$, and radiative corrections depend on the superpartner masses, which 
we assume cannot be too large without reintroducing the hierarchy problem. 
Including the largest radiative corrections at one-loop 
order\footnote{This approximation is subject to significant further 
corrections, which are not necessary for the present argument.} gives:
\beq
M^2_{h^0} \>=\> m_Z^2 \cos^2(2\beta) +
\frac{3 }{4 \pi^2} \sin^2\!\beta \>y_t^2 \biggl [
m_t^2 \, {\rm ln}\left (m_{\tilde t_1} m_{\tilde t_2} / m_t^2 \right )
+ c_{\tilde t}^2 s_{\tilde t}^2 (m_{\tilde t_2}^2 - m_{\tilde t_1}^2)
\, {\rm ln}(m_{\tilde t_2}^2/m_{\tilde t_1}^2) &&
\nonumber   
\\
+ c_{\tilde t}^4 s_{\tilde t}^4 \Bigl \lbrace
(m_{\tilde t_2}^2 - m_{\tilde t_1}^2)^2 - \frac{1}{2}
(m_{\tilde t_2}^4 - m_{\tilde t_1}^4)
\, {\rm ln}(m_{\tilde t_2}^2/m_{\tilde t_1}^2)
\Bigr \rbrace/m_t^2 \biggr ]. && \phantom{xx}
\label{hradcorrmix}
\eeq
where $c_{\tilde t}$ and $s_{\tilde t}$ are the cosine and sine of a 
top-squark mixing angle, $m_{\tilde t_{1,2}}$ are the top-squark mass 
eigenvalues, $y_t$ and $m_t$ are the top-quark Yukawa coupling and mass, 
and $\tan\beta = v_u/v_d$ is the ratio of Higgs vacuum expectation values, 
and for simplicity the Higgs sector is treated in a decoupling 
approximation with ${h^0}$ much lighter than the other Higgs bosons $A^0, 
H^0, H^\pm$. (In this paper, I follow the notations and conventions of 
\cite{primer}.) In order to evade the LEP bound, it is clearly helpful to 
have $m_t$ as large as possible, but the experimental central value 
\cite{Tevatrontopmass} has 
fallen recently. It is also required that $\tan\beta$ is not too small. 
For fixed values of the superpartner masses, it follows that an upper 
bound within the approximation of eq.~(\ref{hradcorrmix}) is
\beq
M^2_{h^0} \><\> m_Z^2 \cos^2(2\beta) +
\frac{3 }{4 \pi^2} \sin^2\!\beta \>y_t^2 m_t^2 \left [
{\rm ln}(m_{\tilde t_2}^2 / m_t^2 ) + 3 \right]
\eeq
in the case that the top-squark mixing is adjusted to have the maximum 
positive impact on $M_{h^0}$. In specific model frameworks without 
carefully adjusted top-squark mixing it is typically found that this bound 
is not close to saturated, so while a non-zero top-squark mixing is quite 
useful for satisfying the LEP bounds for a Standard Model-like lightest 
Higgs scalar, it is also usually necessary for $m^2_{\tilde t_2}/m_t^2$ 
to be fairly large.

This is to be contrasted with the condition for electroweak symmetry
breaking, which for $\tan\beta$ not too small takes the form:
\beq
m_Z^2 &=& -2 \left ( |\mu|^2 + m^2_{H_u} \right ) - 
\frac{1}{v_u} \frac{\partial}{\partial v_u} \Delta V
\,+\, {\cal O}(1/\tan^2\!\beta).
\label{eq:mZ}
\eeq
Here $\Delta V$ is the radiative part of the effective potential with 
$v_u$ treated as a real variable in the differentiation, $\mu$ is the 
supersymmetry-preserving Higgs mass parameter, and $m_{H_u}^2$ is the soft 
supersymmetry breaking mass term for the Higgs field that couples to the 
top quark, which must be negative near the electroweak scale. The 
``supersymmetric little hierarchy problem" is that if supersymmetry 
breaking parameters are large enough to make $M_{h^0}$ exceed the LEP 
bounds, then a tuning at the several percent-level (or worse) might seem 
to be needed in eq.~(\ref{eq:mZ}), so that $|\mu|^2$ and $-m^2_{H_u}$ 
nearly cancel. It has been argued that the level of fine tuning required 
can be quantified with various measures, but it is my view that any such 
metrics are inherently and unavoidably subjective, so they will not be 
used here. Although the little hierarchy problem does not admit 
of rigorous judgments, it can and does cause discomfort and 
doubt regarding the likelihood of finding supersymmetric particles in 
present and future collider searches.

There is no sense in which $|\mu|$ is naturally large, in fact it could 
naturally be 0 even in the presence of arbitrary supersymmetry breaking if 
it were not for experimental constraints. The radiative effective 
potential contribution to eq.~(\ref{eq:mZ}) is not negligible, but since 
it is loop-suppressed, it does not imply a drastic fine tuning. Therefore, 
the supersymmetric
little hierarchy problem, if indeed there is one, is implied by the 
fact that $|m^2_{H_u}|$ might be expected to be much larger than $m_Z^2$ 
in models with heavy top squarks. This indeed occurs in popular models 
with few parameters with universal soft supersymmetry breaking terms 
imposed near the scale of apparent gauge coupling unification (the GUT 
scale), hereafter referred to as mSUGRA. However, it has long been 
appreciated that this connection is modified or lost in more general 
models of supersymmetry breaking. In section \ref{sec:compressed}, I will 
review the arguments that suggest that the little hierarchy problem is 
ameliorated in particular by models that predict a smaller gluino mass 
than in unified models.

A further source of tension on the parameter of the MSSM is provided by 
the opportunity of the explaining the cold dark matter by the thermal 
relic density of a neutralino LSP ($\tilde N_1$). Roughly, the 
annihilation rate for neutralinos decreases with increasing supersymmetry 
breaking masses in the absence of special mechanisms dependent on 
particular mass ratios. If the LSP is bino-like, as predicted by many 
mSUGRA models, then the predicted thermal relic abundance is often found 
to be too high\footnote{It is also important that the dark matter need not 
be neutralinos with a thermal relic density.  The LSP might be a gravitino 
or axino, or something else. Or, if the predicted thermal relic abundance 
of neutralino dark matter is too low or too high, it can be enhanced or 
diluted by some non-thermal effect; see for example \cite{nonthermal}. 
However, models that can explain the dark matter without multiplying 
hypotheses should be accorded special interest.} compared to the 
results of WMAP and other experiments \cite{WMAP}-\cite{PDG}. 
The exceptional possibilities have lately been classified qualitatively in 
four main categories, depending on the mechanism most responsible for 
reducing the predicted dark matter density to an acceptable level.

First, in the ``bulk region" of parameter space, there is a relatively 
light neutralino LSP, which pair annihilates by the $t$-channel and 
$u$-channel exchange of sleptons. However, in mSUGRA and similar models, 
this bulk region often predicts that $M_{h^0}$ is too small, or that other 
states should have been detected at LEP or the Fermilab Tevatron 
$p\overline p$ collider, or gives trouble with other indirect constraints.

Second, in the Higgs resonance (or funnel) region, neutralino pairs 
annihilate through the $s$-channel exchange of the pseudo-scalar Higgs 
boson $A^0$ in an $s$-wave final state. Because the relevant coupling is 
proportional to $m_b \tan\beta$, this usually entails large values of 
$\tan\beta$ \cite{Drees:1992am}. (There is also the possibility
of annihilating near the $h^0$ pole \cite{hzeropole}.) 

Third, there is the possibility that the LSP has a significant higgsino 
component, allowing larger neutralino pair annihilation and 
co-annihilation with the heavier neutralinos and charginos, to and through 
weak bosons \cite{Edsjo:1997bg}. This occurs for example in the ``focus 
point" region of parameter space, in which $|\mu|$ is not too large, even 
if the sfermions are very heavy \cite{focuspoint}.

A fourth possibility, the ``sfermion co-annihilation region" of parameter 
space \cite{GriestSeckel}, is obtained if there is a sfermion (typically a 
tau slepton \cite{staucoannihilation} in mSUGRA, but possibly a top squark 
\cite{stopcoannihilationone}-\cite{stopcoannihilationfive}) that happens 
to be slightly heavier than the LSP. A significant density of this 
sfermion will then coexist with the LSP around the freeze-out time, and so 
annihilations involving the sfermion with itself or with the LSP will 
further dilute the number of superpartners and so the eventual dark matter 
density. The co-annihilation region generally requires just the right mass 
difference between the stau or stop quark and the LSP, and so is often 
considered to be fine tuned.

If the LSP is mostly higgsino or wino, then the annihilation of 
superpartners in the early universe is typically too efficient to provide 
for thermal relic density in agreement with WMAP. However, one can always 
adjust the higgsino or wino contents of $\tilde N_1$ to be just right, 
at the expense of some fine tuning. In 
recent years, there have been many studies of the properties of 
neutralino dark matter that follow from 
abandoning the strict boundary conditions of 
mSUGRA models to allow non-universal gaugino masses
\cite{Corsetti:2000yq}-\cite{Bae:2007pa} or 
scalar masses \cite{Berezinsky:1995cj}-\cite{Evans:2006sj}
at the GUT scale.
By increasing the wino or higgsino content of the neutralino,
one can increase the cross-section for annihilations and co-annihilations
to weak bosons, and those mediated by the $Z$ boson and $h^0$ and $A^0$ in
the $s$-channel.

In section \ref{sec:dark} of this paper, I will study a different 
possibility with rather distinctive properties, namely the possibility 
that the LSP is mostly bino-like, but pair annihilates efficiently to 
top-antitop quark pairs due predominantly to the exchange of light 
top squarks. 
In the models discussed in section \ref{sec:compressed} (unlike in mSUGRA 
and similar models) this mechanism turns out to give a thermal relic dark 
matter density in agreement with the WMAP measurements for a wide range of 
input parameters, much more than in the stop co-annihilation (or stau 
co-annihilation) regions to which it is continuously connected. This 
scenario also has important implications for collider searches at the 
Fermilab Tevatron, CERN Large Hadron Collider (LHC), and a future linear 
collider, discussed briefly in section \ref{sec:colliders}. Section 
\ref{sec:outlook} contains some concluding remarks.

\section{Compressed supersymmetry}\label{sec:compressed}
\setcounter{equation}{0}
\setcounter{footnote}{1}

In this section, I review the argument that a suppression of the gluino 
mass parameter ameliorates the little hierarchy problem in supersymmetry. 
(This has been observed in various papers; a particularly clear and early 
explanation was given in ref.~\cite{KaneKing}.)

As noted in the Introduction, the issue is essentially to explain why the 
running parameter $-m_{H_u}^2$ should be small and positive near the 
electroweak scale, in the same theory that allows large positive 
corrections to $M_{h^0}^2$. The parameter $|\mu|^2$, which relies on a 
different sector of the theory, can then be chosen without too much fine 
tuning to give the right $m_Z^2$ in eq.~(\ref{eq:mZ}). In terms of the 
MSSM soft supersymmetry breaking parameters at the apparent GUT scale, one 
finds:
\beq
-m_{H_u}^2 &=&
1.92 \barM_3^2
+ 0.16 \barM_2 \barM_3
-0.21 \barM_2^2
-0.33 \barM_3 \barA_t 
-0.074 \barM_2 \barA_t
+ 0.11 \barA_t^2 
\nonumber \\ &&
+0.024 \barM_1 \barM_3
 +0.006 \barM_1 \barM_2
- 0.006 \barM_1^2
 - 0.012 \barM_1 \barA_t 
+ 0.002 \barM_3 \barA_b 
\nonumber \\ &&
- 0.63 \barm^2_{H_u} + 0.36 \barm^2_{Q_3} +0.28 \barm^2_{u_3}
-0.027 \barm^2_{H_d} +0.025 \barm^2_{d_3} - 0.026 \barm^2_{L_3} 
\nonumber \\ &&
+ 0.026 \barm^2_{e_3}
+ 0.05 \barm^2_{Q_1} -0.11 \barm^2_{u_1}  +0.05 \barm^2_{d_1} 
- 0.05 \barm^2_{L_1} + 0.05 \barm^2_{e_1} 
\label{eq:mHu}
\eeq
Here, the hats on the parameters on the right-hand side denote that they 
are inputs at the apparent GUT scale, while $m^2_{H_u}$ on
the left-hand side denotes the 
result at the renormalization
scale $Q=400$ GeV (where the corrections due to the effective 
potential are presumed moderate), using $\tan\beta=10$ and the 
SPS1a$'$ benchmark point \cite{sillybenchmarkpoint} values for the Yukawa 
and gauge couplings and unification scale,
and using two-loop renormalization group equations \cite{twoloopRGEs}. 
The input parameters consist of independent gaugino masses $\barM_1$, 
$\barM_2$, $\barM_3$, scalar trilinear coupling parameters $\barA_t$, 
$\barA_b$, $\barA_\tau$, and scalar squared masses for the Higgs bosons, 
third family sfermions, and first family sfermions (with each second 
family sfermion assumed degenerate with the first family counterpart 
having the same quantum numbers). 
Some contributions with very small coefficients have 
been omitted from eq.~(\ref{eq:mHu}). 
The reason for applying boundary 
conditions at the GUT mass is that the apparent unification of couplings 
provides some justification that it is meaningful to extrapolate running 
parameters up to that scale.

In the so-called mSUGRA framework, the input parameters are usually taken 
to obey the much stronger conditions:
\beq
&&
\barM_1 = \barM_2 = \barM_3 = m_{1/2},
\\
&&
\barA_t = \barA_b = \barA_\tau = A_0,
\\
&&
\barm^2_{\phi} = m_0^2
\eeq
for all scalars $\phi = H_u, H_d, Q_i, u_i, d_i, L_i, e_i$, with family 
index $i=1,2,3$. It is then clear that the largest contribution to 
$-m^2_{H_u}$ at the weak scale is due to the input gluino mass $\barM_3$; 
furthermore, there is a significant cancellation between the scalar 
contributions within the mSUGRA framework.

Generalizing the input parameters can provide a relative reduction in 
$-m_{H_u}^2$, therefore lowering both the predicted value of $|\mu|^2$ and 
the cancellation needed to obtain the observed value of $m_Z$. From consideration of the first five terms on the right-hand side, one 
learns that small values of $|\mu|$ result from (roughly) 
$\barM_3 \approx 
0.29 \barM_2 + 0.13 \barA_t$ or 
$\barM_3 \approx -0.38 \barM_2 + 0.04 \barA_t$, provided that $\barM_1$, 
$\barA_t$, $\barm_{H_u}^2$ etc.~are not too large.
A complete cancellation is 
actually not desirable from our present point of view, since a mostly 
higgsino-like neutralino has too small a thermal relic density, and 
$M_{h^0}$ often comes out too small. There are many types of models
already in the literature 
that can predict a small ratio of $\barM_3/\barM_2$. 
The scenario for dark matter to be studied in the next section 
does not depend crucially on which framework is used, 
but for concreteness I will review one here.

In the usual mSUGRA framework, one assumes that the gaugino masses are all 
the same; in $SU(5)$ GUT language this corresponds to a supersymmetry 
breaking $F$-term in a singlet of $SU(5)$ [or $SO(10)$]. More generally, 
one can consider non-universal gaugino masses arising from an $F$ term VEV 
in arbitrary linear combinations of the symmetric product of two adjoint 
representations of the GUT group that contain a Standard Model singlet 
\cite{Ellis:1985jn}-\cite{Anderson:1999ui}.
For $SU(5)$:
\beq
({\bf 24}\times {\bf 24})_S = {\bf 1} + {\bf 24} + {\bf 75} + {\bf 200}.
\eeq 
The resulting gaugino mass terms 
have the form
\beq 
{\cal L} = -\sum_{R} \frac{\langle F_{R} \rangle}{2M_P}
\sum_{n} c^{(n)}_{R} \lambda_n \lambda_n + {\rm c.c.}
\eeq
where the coefficients $c^{(n)}_{R}$ 
(with $n=1,2,3$ for bino, wino, and gluino 
respectively, and $R = {\bf 1} + {\bf 24} + {\bf 75} + {\bf 200}$) 
are determined by group theory, leading to the 
parameterization: 
\beq
\barM_1 &=& m_{1/2} (1 + \ctf + 5 \csf + 10 \cth),
\label{eq:M1fromadj}
\\
\barM_2 &=& m_{1/2} (1 + 3 \ctf -3 \csf + 2 \cth),
\\
\barM_3 &=& m_{1/2} (1 -2 \ctf - \csf + \cth).
\label{eq:M3fromadj}
\eeq
The special case $\ctf = \csf = \cth$ recovers the mSUGRA model. In
eqs.~(\ref{eq:M1fromadj})-(\ref{eq:M3fromadj}), I have assumed that
there is at least some $SU(5)$ singlet component to the $F$ term,
although this is not strictly necessary. Note that this parameterization 
is already general enough to fit any observed gaugino mass hierarchy,
simply because it contains three linearly independent contributions.

One particularly simple way to achieve a ratio $\barM_3/\barM_2 \sim 
1/3$ is to choose $\ctf \sim 0.22$, with $\csf = \cth = 0$. This is the 
inspiration for the model space studied in the next section, although it 
cannot be overemphasized that there are many other reasonable ways to 
achieve such a ratio. One point in favor of this type of model
is that in GUT theories like $SU(5)$, there is a chiral superfield
in the ${\bf 24}$ (adjoint) representation anyway; once the
scalar component acquires
a VEV, it is actually unnatural for the $F$ term to not develop a VEV
as well. Moreover, this can make the theory consistent with
proton decay requirements and help to obtain precise
gauge coupling unification \cite{Huitu:1999eh,TobeWells}.

As evidenced by the special case of eq.~(\ref{eq:mHu}), soft supersymmetry 
breaking mass parameters at the weak scale are substantially driven by the 
gaugino mass parameters through large logarithmic effects that are 
summarized in the renormalization group. In mSUGRA models, this effect 
typically causes squarks and the gluino to be much heavier than the 
superpartners that do not have $SU(3)_C$ interactions, except when $m_0$ 
is very large. In the case of a small ratio $\barM_3/\barM_2$ motivated by 
a solution to the little hierarchy problem, however, the resulting mass 
spectrum will be ``compressed" in comparison to mSUGRA, with a smaller 
ratio of the masses of the heaviest and lightest superpartners. Two 
aspects of this that will be important in the next section are that it 
becomes much more likely that the lighter top squark can be the 
next-to-lightest supersymmetric particle (NLSP), and that the LSP is 
rather heavy.

It will be assumed here 
that the trilinear scalar couplings are sizable and negative 
at the GUT scale (in the convention of \cite{primer}). This can be
motivated as the 
effect of strong
 renormalization group running between the GUT scale and 
the Planck scale, which would produce both negative scalar trilinear 
couplings and positive scalar squared masses, and often
prefers positive $\mu$ 
\cite{muispositive} (independently of low-energy $(g-2)_\mu$ or $b 
\rightarrow s \gamma$ considerations), when the running is dominated by 
positive gaugino masses. 
It is worth noting that in many viable supersymmetric GUT theories,
the naive running of the gauge couplings above the unification scale quickly
becomes non-perturbative. For example, in the minimal missing partner
supersymmetric $SU(5)$ theory
\cite{missingpartner}, the two-loop gauge beta function has a Landau
pole, and the three- and four-loop beta functions appear to have
strongly-coupled 
UV-stable fixed points \cite{UVstable}. 
While the breakdown of perturbation theory
renders such calculations untrustworthy in detail, this suggests that
gaugino mass 
dominance could eliminate the problem of supersymmetric flavor-violation,
while giving essentially arbitrary flavor-preserving soft supersymmetry
breaking terms at the GUT scale.
As a simplistic assumption 
made only for convenience, the scalar trilinear 
couplings $\barA_t$, $\barA_b$, $\barA_\tau$ will be taken to be unified
in the models below; the parameter $\barA_t$ has the most direct 
importance in most cases.

Likewise I will assume, as a matter of convenience, 
that at the GUT scale all scalars have 
a common squared mass $m_0^2$ as in mSUGRA. While it is clearly 
worthwhile to 
consider scalar mass non-universality at the GUT scale, 
I expect that the results obtained here will be realized at least 
qualitatively in a variety of different schemes without universal scalar 
squared masses imposed at the GUT scale. 

\section{Dark matter density and pair annihilation to 
top quarks}\label{sec:dark}
\setcounter{equation}{0}
\setcounter{footnote}{1}

In order to explain a large value of $M_{h^0}$ in models of compressed 
supersymmetry, it is favored that the mass spectrum is compressed up, 
rather than down. This means that the LSP will have to be heavier than 
usually found in the mSUGRA ``bulk" region for dark matter.

It has been suggested in models of this type with small
$|M_3/M_2|$ that the thermal relic 
abundance of dark matter can be explained by an enhanced Higgsino component of 
$\tilde N_1$, leading to enhanced annihilations $\tilde N_1 \tilde N_1 
\rightarrow W^+W^-$ or $ZZ$ \cite{Bertin:2002sq,Baer:2006dz}, 
or by $s$-channel annihilation through the 
pseudo-scalar Higgs $A^0$ near resonance 
\cite{Bertin:2002sq,Belanger:2004ag,Mambrini:2005cp}, or 
by co-annihilations with heavier neutralinos and charginos
\cite{Belanger:2004ag,Baer:2006dz}, 
or by $s$-channel annihilations to $t\overline t$ 
through the $Z$ boson \cite{Bertin:2002sq,Mambrini:2005cp}.

In this paper, I will consider instead the case that the thermal relic 
density is suppressed primarily by $\tilde N_1$ pair annihilation to top 
quark-antiquark pairs, mediated mostly by top-squark exchange. As 
mentioned in the previous section, it is not difficult in compressed 
supersymmetric models to obtain a top squark NLSP. This is in 
contradistinction to mSUGRA and similar models, where achieving the 
required suppression in $\Omega_{\rm DM} h^2$ from top-squark exchange 
requires that $|A_0|$ is very large in absolute terms and must be rather 
finely adjusted so that $\tilde t_1$ is not much heavier than $\tilde N_1$ 
(see for example 
refs.~\cite{stopcoannihilationtwo,stopcoannihilationthree,%
stopcoannihilationfive}). In the points 
in mSUGRA parameter space where this can occur, $\tilde t_1 \tilde t_1$ 
and $\tilde N_1 \tilde t_1$ co-annihilations are also generally very 
important (unlike here). Compressed supersymmetry models with 
small\footnote{Hats will be omitted from GUT scale input parameters 
throughout this section.} $|M_3/M_2|$ at the GUT scale have the crucial 
distinction that achieving comparable $m_{\tilde t_1}$ and $m_{\tilde 
N_1}$ is much easier, and requires smaller values of $|A_t|$ in absolute 
terms, and admits a wider range of $A_t$.
 
\begin{figure}
\begin{picture}(120,55)(0,0)
\SetWidth{0.85}   
\Line(0,0)(100,0)
\Line(0,50)(100,50)
\DashLine(50,0)(50,50){5}
\rText(-13,8)[][]{$\widetilde N_1$}   
\rText(-13,45)[][]{$\widetilde N_1$}  
\rText(57,25)[][]{$\tilde t_{1,2}$}
\rText(101,48)[][]{$ t$}
\rText(101,6)[][]{$\overline t$}
\end{picture}
\hspace{2.7cm}
\begin{picture}(120,55)(0,0)
\SetWidth{0.85}   
\Line(0,0)(22,22)
\Line(28,28)(50,50)
\Line(0,50)(50,0)
\Line(50,0)(100,0)
\Line(50,50)(100,50)
\DashLine(50,50)(50,0){5}
\rText(-13,9)[][]{$\widetilde N_1$}   
\rText(-13,44)[][]{$\widetilde N_1$}  
\rText(57,25)[][]{$\tilde t_{1,2}$}
\rText(101,48)[][]{$ t$}
\rText(101,6)[][]{$\overline t$}
\end{picture}

\vspace{0.9cm}

\begin{picture}(120,60)(0,0)
\SetWidth{0.85}   
\Line(0,50)(33,25)  
\Line(77,25)(110,50)
\Photon(33,25)(77,25){2.1}{4.5}
\Line(0,0)(33,25)  
\Line(77,25)(110,0)
\rText(-12,8)[][]{$\tilde N_1$}  
\rText(-12,47)[][]{$\tilde N_1$}
\rText(51.9,34)[][]{$Z$} 
\rText(110,47.5)[][]{$t$}
\rText(109.5,3)[][]{$\overline{t}$}
\end{picture}
\hspace{2.5cm}
\begin{picture}(120,60)(0,0)
\SetWidth{0.85}
\Line(0,0)(30,25)
\Line(0,50)(30,25)
\DashLine(30,25)(90,25){5}
\Line(120,0)(90,25)
\Line(120,50)(90,25)
\rText(-12,8)[][]{$\tilde N_1$}
\rText(-12,47)[][]{$\tilde N_1$}
\rText(55.3,32.8)[][]{$A^0, h^0, H^0$}
\rText(120.5,48)[][]{$t$}
\rText(121,3)[][]{$\bar t$}
\end{picture}
\vspace{0.1cm}
\caption{Contributions to the annihilation of neutralino
dark matter LSP pairs into top quark-antiquark pairs, from top squark, 
$Z$ boson, and Higgs boson exchange. 
\label{fig:annihilation}}
\end{figure}
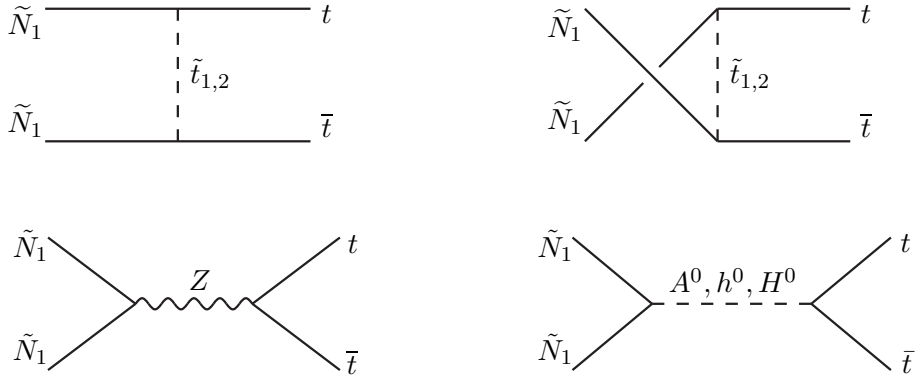

The tree-level Feynman 
diagrams that contribute to the process $\tilde N_1 \tilde N_1 \rightarrow 
t \overline t$ are shown in Figure \ref{fig:annihilation}.
In order to obtain $\Omega_{\rm DM} h^2$ compatible with WMAP by this
mechanism (without undue fine tuning), it is necessary that:
\beq
&&
m_t \,<\, m_{\tilde N_1} \,\lsim\, m_t + 100\>{\rm GeV},
\\
&&m_{\tilde N_1} + 25\>{\rm GeV} \,\lsim \,
m_{\tilde t_1} 
\,\lsim \,
m_{\tilde N_1} + 100\>{\rm GeV}.
\label{eq:stopbounds}
\eeq
The first inequality in eq.~(\ref{eq:stopbounds}) is the approximate 
requirement that the relic density not be suppressed too much by 
top-squark co-annihilations. The upper bounds here are also necessarily 
fuzzy, because of the connection to the thin co-annihilation region. For 
models satisfying these criteria, the top-squark exchange is most 
important for bino-like $\tilde N_1$ and $\tilde t_1$ with a high $\tilde 
t_R$ component. As one increases the small Higgsino component of $\tilde 
N_1$ (by lowering $|\mu|$), the contribution from the $\tilde t_1$ 
exchange diagrams becomes
enhanced, due to the top Yukawa coupling. In the 
models to be considered below, the $s$-channel
$Z$ exchange diagram is subdominant but 
not negligible; using the analytic formulas provided in 
\cite{Drees:1992am}, one can show that the most important effect is a 
significant destructive interference with the dominant top-squark exchange 
diagram amplitude.

For a more detailed study, I have used the program {\tt micrOMEGAs 2.0.1} 
\cite{micrOMEGAs} to evaluate the thermal relic abundance of dark matter 
for supersymmetric models generated using {\tt SOFTSUSY 2.0.11} 
\cite{softsusy} (and checked for approximate agreement with {\tt SuSpect}
\cite{suspect}). In the following, I consider a rather conservative 
thermal relic density constraint:
\beq
0.09 < \Omega_{\rm DM} h^2 < 0.13,
\label{eq:WMAPconstraint}
\eeq 
and impose a Higgs mass constraint:
\beq
M_{h^0} > 113\>{\rm GeV}.
\label{eq:Mhbound}
\eeq
This is slightly lower than the LEP bound for Standard Model-like Higgs 
scalars, justified by the significant uncertainties involved in the 
theoretical Higgs mass calculation. In addition, I adopt the slightly 
optimistic value of $m_t = 175$ GeV rather than the somewhat lower latest 
combined
central value $m_t = 171.4 \pm 1.8\,{\rm(syst.)} \pm 1.2 \,{\rm(stat.)}$ 
GeV from the Tevatron \cite{Tevatrontopmass}. In each model, I require 
that the LSP is a neutralino. Then all limits on supersymmetric particles 
from LEP turn out to be automatically satisfied. No constraint from the 
anomalous magnetic moment of the muon is applied, since for all models 
considered here, the predicted value is actually closer to the 
experimental central value(s) from the BNL E821 experiment 
\cite{BNLmuongminustwo} than the Standard Model prediction is (but not by 
a very statistically significant amount). I do not impose any bound coming 
from $b \rightarrow s \gamma$, since the measurement can be easily 
accommodated \cite{bsgammaisbogus} by introducing some extra small 
GUT-scale flavor violation in the supersymmetry-breaking parameters, in a 
way that would not affect the rest of the model in any appreciable way.

\begin{figure*}[!p]
\vspace{-0.2cm}
\centering
\mbox{\includegraphics[width=11cm]{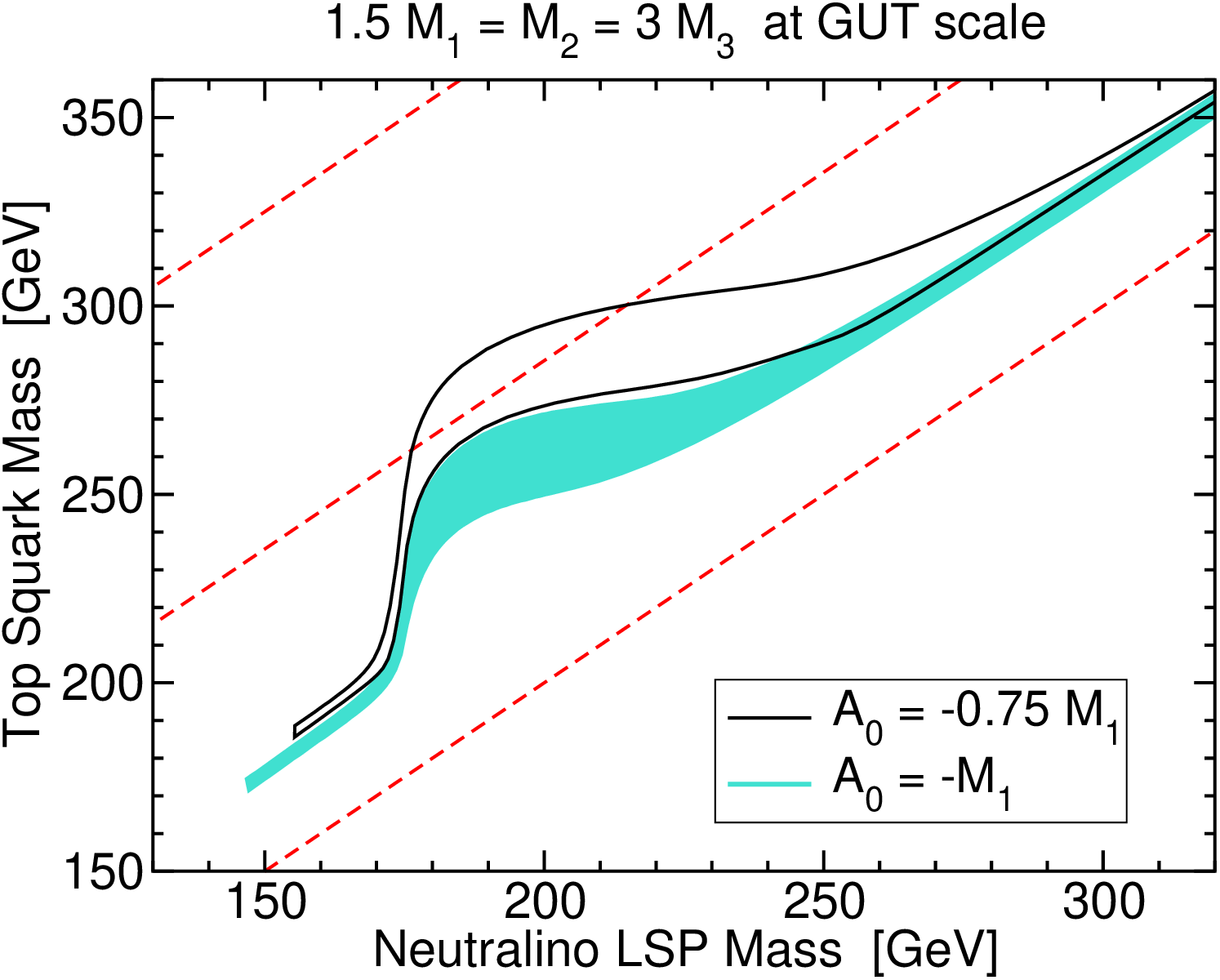}}
\caption{\label{fig:LSPstop}
Allowed regions in the $m_{\tilde N_1}$, $m_{\tilde t_1}$ plane
that predict a thermal relic abundance of neutralino LSP
dark matter $0.09 < \Omega_{\rm DM} h^2 < 0.13$ and satisfy other 
constraints
given in the text. 
The GUT-scale parameters satisfy
$1.5 M_1 = M_2 = 3 M_3$, with variable $m_0$, and
$A_0 = -0.75 M_1$ (region outlined in black) and $A_0 = -M_1$
(shaded region). Also, $\tan\beta = 10$ and $\mu>0$.
The lowest dashed line is $m_{\tilde t_1} = m_{\tilde N_1}$.
Below the upper dashed line, the 
decay $\tilde t_1 \rightarrow t \tilde N_1 $ is 
forbidden.
Below the middle dashed line, the three-body
decay $\tilde t_1 \rightarrow W b \tilde N_1 $ is also forbidden.}
%
\vspace{0.8cm}
\mbox{\includegraphics[width=11.3cm]{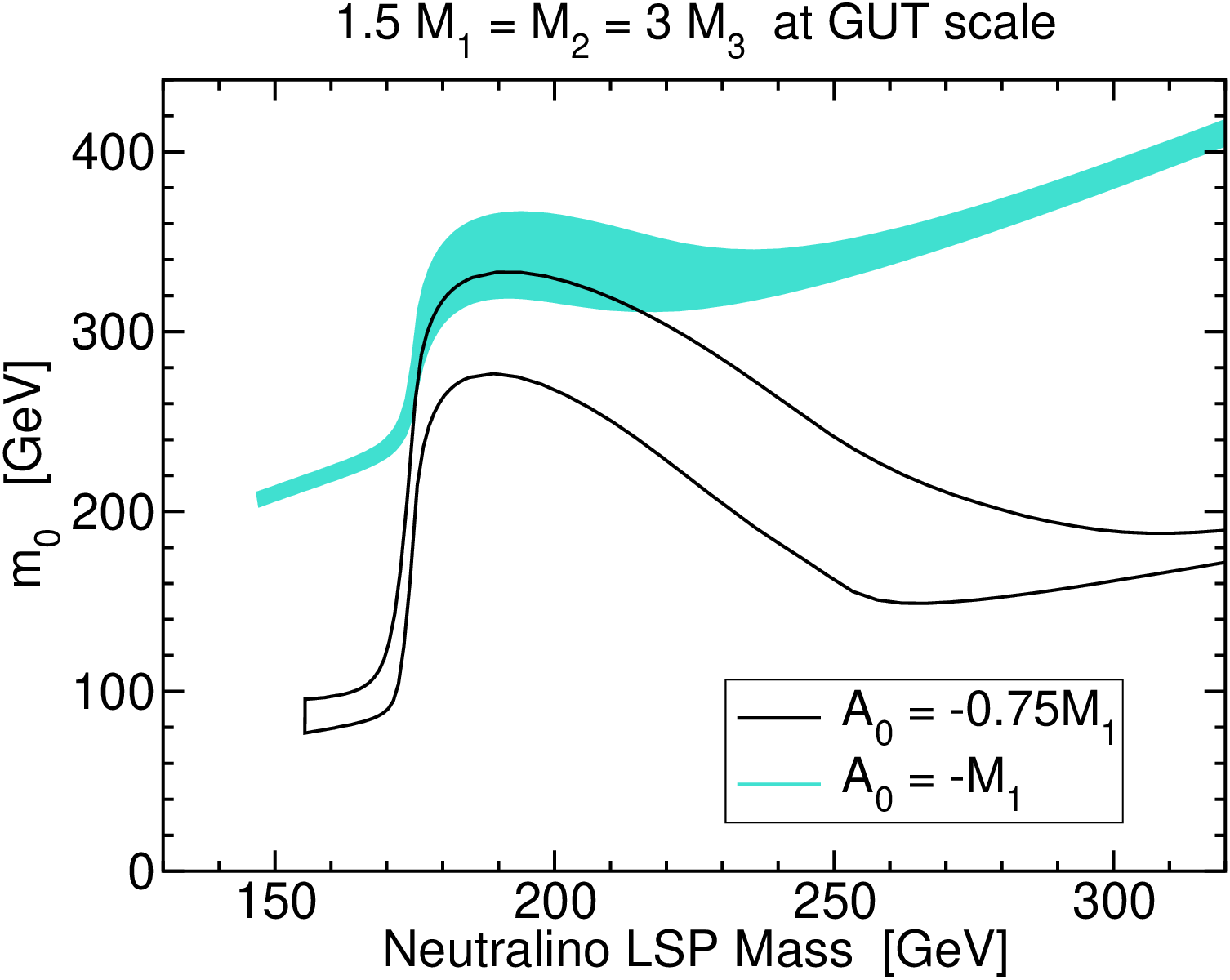}}
\caption{\label{fig:LSPmzero}
The allowed regions depicted in Figure 1 arise from the values of
$m_0$ shown here.}
\end{figure*}

Results for a typical two-parameter model space are shown in Figure 
\ref{fig:LSPstop}. Here I consider models with boundary conditions at the 
GUT scale:
\beq
1.5 M_1 = M_2 = 3 M_3,
\eeq
with $M_1$ and $m_0$ allowed to vary independently, 
$\tan\beta = 10$ and $\mu>0$, 
and two values for the ratio $A_0/M_1 = -0.75$ (outlined region) and 
$A_0/M_1 = -1$ (shaded region). The allowed regions are cut off on the 
lower left by the Higgs mass bound constraint eq.~(\ref{eq:Mhbound}). The 
upward bulges in the regions are the places where top-squark exchange 
plays the dominant role in $\tilde N_1 \tilde N_1 \rightarrow t \overline 
t$, which in turn is the most important annihilation process for the dark 
matter. Typically, in the bulge regions 
$\tilde N_1 \tilde N_1 \rightarrow t \overline t$
accounts for 70\% to 90\% of the contributions to $1/\Omega_{\rm DM} h^2$.
Each of these regions is continuously connected to much narrower 
co-annihilation regions on either side. For $A_0/M_1 = -1$, stop 
co-annihilation is the dominant effect in these thin strips, but for 
$A_0/M_1 = -0.75$, stau co-annihilations are also important there.

For smaller values of $-A_0/M_1$, not shown here, the $\tilde N_1 \tilde 
N_1 \rightarrow t \overline t$ bulge region still exists, but continuously 
connects instead to a thin stau co-annihilation strip.

As can be seen from Figure \ref{fig:LSPstop}, the smaller $-A_0$ case 
requires a larger mass difference $m_{\tilde t_1} - m_{\tilde N_1}$ in the 
allowed bulge region. This can be traced to the fact that $\mu$ decreases 
as $-A_0$ decreases [see eqs.~(\ref{eq:mZ}) and (\ref{eq:mHu})], slightly 
enhancing the small Higgsino component of the LSP, which substantially 
increases the top-squark exchange amplitude contributions to the 
annihilation cross section as mentioned above.

The $s$-channel Higgs scalar annihilation diagrams shown in Figure 
\ref{fig:annihilation} play only a small role in the LSP pair annihilation 
in these models. This is because in all cases $m_{\tilde N_1}$ is well 
below the resonance point $m_{A^0}/2 \,\approx \, m_{H^0}/2$, and well 
above the resonance point $m_{h^0}/2$.

Also shown in Figure \ref{fig:LSPstop} are the critical lines that govern 
the decay of $\tilde t_1$. The lowest dashed line indicates the region 
allowed by the requirement that $\tilde N_1$ is the LSP. In all cases, the 
decay $\tilde t_1 \rightarrow t \tilde N_1$ is kinematically closed, as 
indicated by the upper dashed line. Above the middle dashed line, the 
decay
\beq
\tilde t_1 \rightarrow Wb\tilde N_1
\eeq
is kinematically open, and should dominate. Below that line, 
the flavor-violating 2-body decay
\beq
\tilde t_1 \rightarrow c \tilde N_1
\eeq
is expected to win \cite{Hikasa:1987db} over the four-body decays
$
\tilde t_1 \rightarrow q \overline q' b\tilde N_1
$ 
and
$
\tilde t_1 \rightarrow \ell \nu b \tilde N_1
$.
The charginos ($\tilde C_i$) and sleptons are heavier in the models
shown, so they cannot appear in $\tilde t_1$ decays.

The relative thickness of the allowed regions
cannot be regarded as a direct measure of fine-tuning, even 
subjectively. In fact, a perfectly accurate determination of $\Omega_{\rm 
DM} h^2$ would make the allowed regions arbitrarily thin, up to model 
assumption and theoretical errors and input parameter 
uncertainties.\footnote{Moreover, one could 
regard the entire parameter space between the indicated regions and the 
$m_{\tilde t_1}=m_{\tilde N_1}$ line as allowed, in the sense that the 
thermal relic abundance would be less than or equal to the observed value. 
Then something else, for example axions, would make up the remaining dark 
matter.}
(The 
Planck satellite mission experiment \cite{Planck} should indeed 
significantly improve the determination.) However, it seems clear that the 
observed dark matter density is more naturally accommodated in the $\tilde 
N_1 \tilde N_1 \rightarrow t \overline t$ bulge regions, since a larger 
range of $m_0$ values (for each fixed $M_1$) lie within the range 
specified by eq.~(\ref{eq:WMAPconstraint}). This is illustrated for the 
same models in Figure \ref{fig:LSPmzero}. Notably, all of the soft 
supersymmetry breaking parameters are less than the gaugino masses $M_1$ 
and $M_2$; this includes the holomorphic term $b = B\mu$, which is of 
order (250 GeV)$^2$ in the bulge region of these models. 

\begin{figure*}[tpb] \centering \mbox{\includegraphics[width=12cm]{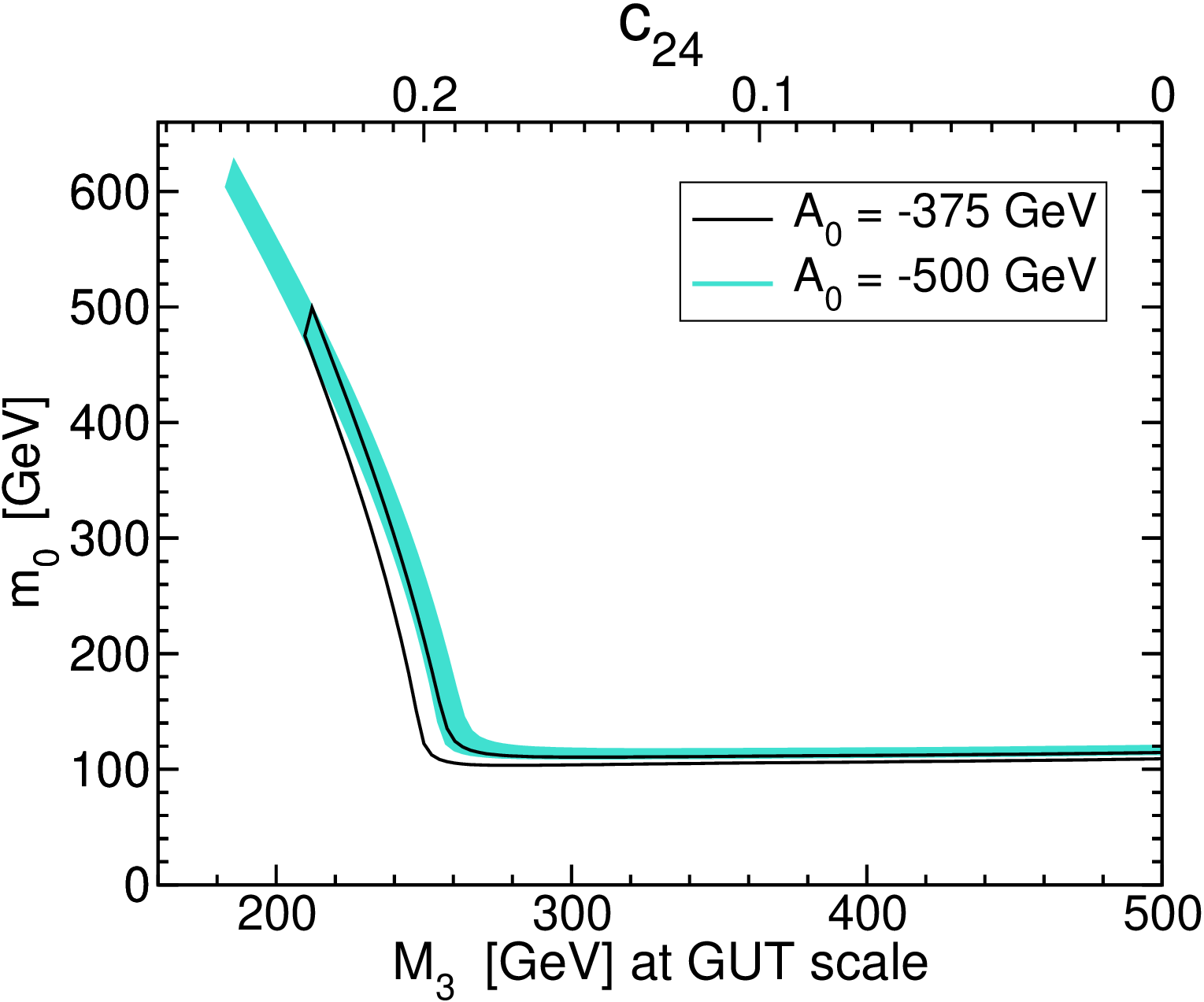}} 
\caption{\label{fig:c24m0} Allowed regions that predict a thermal relic 
abundance of neutralino LSP dark matter $0.09 < \Omega_{\rm DM} h^2 < 
0.13$ and satisfy other constraints given in the text. At the GUT scale, 
the bino mass parameter $M_1 = 500$ GeV is fixed, and the wino and gluino 
mass parameters vary while obeying 
eqs.~(\ref{eq:M1fromadj})-(\ref{eq:M3fromadj})
with $C_{75} = C_{200} = 0$. The horizontal direction 
is parameterized by $M_3$ at the GUT scale (lower horizontal axis) or 
equivalently by $\ctf$ (upper horizontal axis). The vertical axis is the 
common scalar mass $m_0$. The region outlined in black has 
$A_0 = -375$ GeV and the shaded region has $A_0 = -500$ GeV, with 
$\tan\beta=10$ and $\mu > 0$ in each case. The very thin, nearly 
horizontal regions with $m_0$ near 110 GeV feature co-annihilation of 
sleptons and the LSP. In the thicker sloping areas on the left, the 
dominant contribution to $1/\Omega_{\rm DM} h^2$ is $\tilde N_1 \tilde N_1 
\rightarrow t \overline t$, mostly due to the $\tilde t_1$ exchange 
diagram amplitudes.} \end{figure*} 

Another handle on the $\tilde N_1 \tilde N_1 \rightarrow t \overline t$ 
annihilation scenario is provided by Figure \ref{fig:c24m0}, which shows 
dark matter allowed regions in the plane of $m_0$ vs. $M_3$, for models 
with fixed $M_1 = 500$ GeV at the GUT scale
(so that the LSP mass is approximately $200$ 
GeV) and obeying the boundary condition of 
eqs.~(\ref{eq:M1fromadj})-(\ref{eq:M3fromadj}) with $\ctf$ varying and 
$\csf = \cth = 0$. I again require $\mu>0$ and $\tan\beta=10$, and the 
allowed regions are shown for $A_0 = -M_1$ and $A_0 = -0.75 M_1$. The thin 
horizontal regions achieve the observed dark matter density by 
co-annihilations of sleptons and the LSP; as is well-known, this requires 
a rather precise adjustment of the slepton squared masses. For $\ctf \gsim 
0.19$, or equivalently $M_3 \lsim 260$ GeV, the $\tilde N_1 \tilde N_1 
\rightarrow t \overline t$ annihilation scenario takes over, leading to 
the thicker, sloping allowed regions. They are cut off on the left by the 
imposed Higgs mass constraint eq.~(\ref{eq:Mhbound}).

\begin{figure*}[tpb]
\centering
\mbox{\includegraphics[width=11cm]{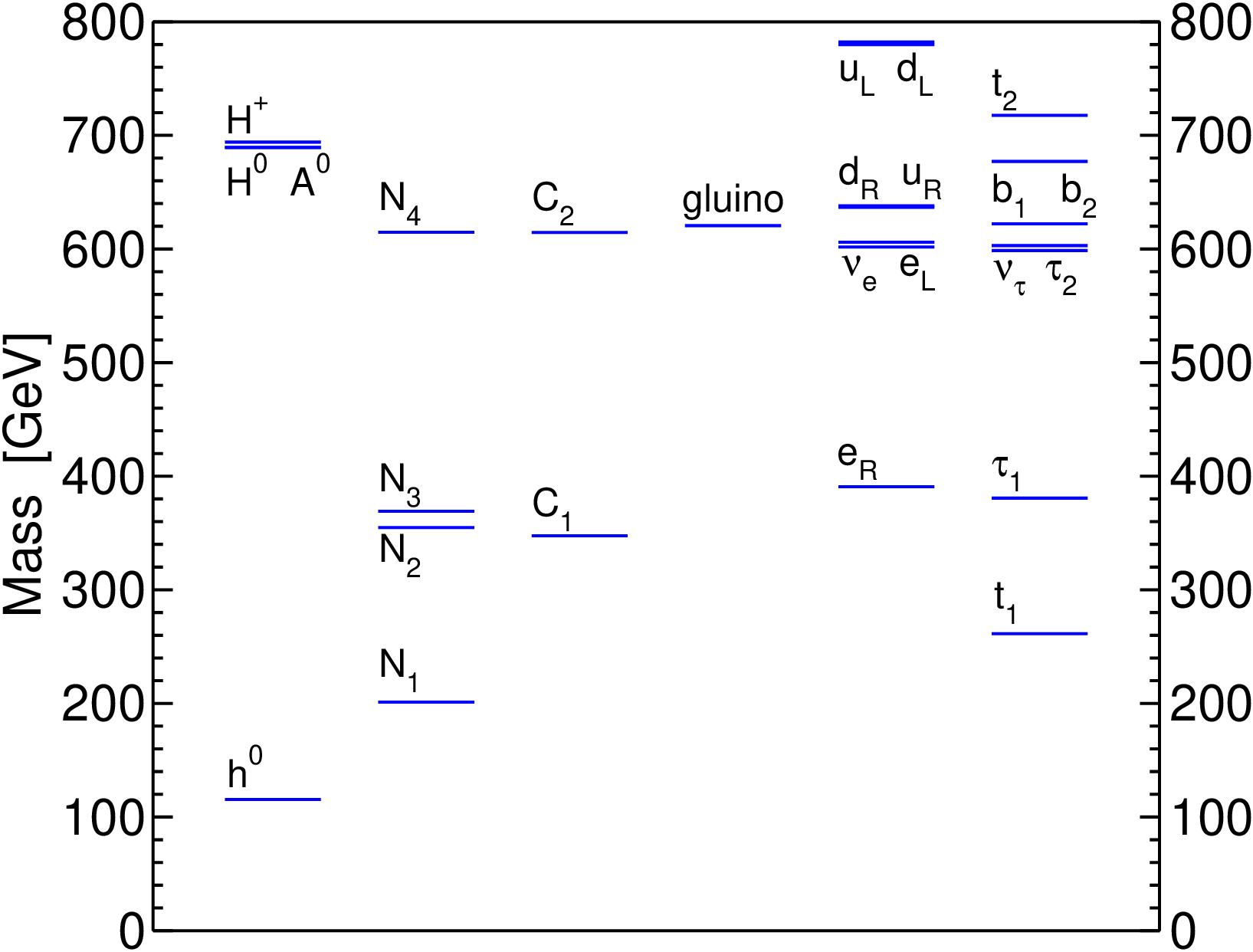}}
\caption{\label{fig:spectrum}
A typical sample ``compressed" Higgs and superpartner 
mass spectrum with $\Omega_{\rm DM} h^2 = 0.11$ brought
about by $\tilde N_1 \tilde N_1 \rightarrow t \overline t$
through $\tilde t_1$ exchange. 
The GUT scale parameters of the model
are $M_{1,2,3} = 500, 750, 250$, $A_0 = -500$, and $m_0 = 342$ GeV,
with $\tan\beta= 10$ and $\mu>0$ at the weak scale. The ratio of the
largest superpartner mass to the smallest is less than 4. An
unfortunate feature, quite common to this scenario for dark matter, 
is that no visible superpartners would be
within reach of a linear collider with $\sqrt{s} = 500$ GeV.}
\end{figure*}

The distinctive features of the $\tilde N_1 \tilde N_1 \rightarrow t 
\overline t$ annihilation scenario for dark matter in compressed 
supersymmetry are illustrated in the superpartner spectrum for a typical 
model point shown in Figure \ref{fig:spectrum}, with $M_1 = 500$ GeV and 
$m_0 = 342$ GeV in order to give $\Omega_{\rm DM} h^2 = 0.11$.
In this model, $\tilde N_1 \tilde N_1 \rightarrow t 
\overline t$ contributes about 89\% to $1/\Omega_{\rm DM} h^2$.
The amplitude from $\tilde t_1$ exchange is largest, with an amplitude
from $Z$ exchange about $0.3$ times as big in the velocity-independent
part of the ${}^1S_0$ channel, with destructive interference.
The superpartner mass spectrum shows compression compared to mSUGRA 
models, with the ratio of masses of the largest superpartners (nearly 
degenerate $\tilde u_L$, $\tilde c_L$ and $\tilde d_L$, $\tilde s_L$) to 
the LSP being less than 4, with all superpartners between 200 GeV and 800 
GeV. The NSLP is $\tilde t_1$. The lightest chargino $\tilde C_1$ and the 
neutralinos $\tilde N_2$ and $\tilde N_3$ are higgsino-like; this is a 
consequence of $\mu$ being not too large as discussed in section 
\ref{sec:compressed}. Another consequence of the choice of a relatively 
large wino mass to ameliorate the little hierarchy problem is that the 
wino-like states $\tilde N_4$ and $\tilde C_2$ are comparatively heavy, 
just below the gluino mass, and there is a wide split between left-handed 
squarks and sleptons and their right-handed counterparts.

\section{Implications for colliders}\label{sec:colliders}
\setcounter{equation}{0}
\setcounter{footnote}{1}

In this section I will make some brief remarks on the implications of the 
scenario outlined above for collider searches. Figure \ref{fig:spectrum} 
shows a typical model of this type, with the characteristic feature that 
the decay $\tilde t_1 \rightarrow c \tilde N_1$ has a 100\% branching 
fraction. For this section, 
I will use this as a benchmark, with the important caveat that 
search strategies will be qualitatively different if the decay $\tilde t_1 
\rightarrow Wb \tilde N_1$ is open. Other notable decays, with approximate 
branching fractions computed by {\tt ISAJET 7.74} \cite{ISAJET}, are:
\beq
\tilde g \rightarrow 
\Biggl \{ \begin{array}{l}
t \,\tilde t_1^*\qquad(\sim 50\%)
\\
\overline t \,\tilde t_1\qquad(\sim 50\%)
\end{array}\Biggr.
\label{eq:gluinodecay}
\eeq
for the gluino, and
\beq
&&
\tilde N_2 \rightarrow 
\Biggl \{ \begin{array}{ll} 
\tilde N_1 h\quad &(\sim 90\%)
\\
\tilde N_1 Z\quad &(\sim 10\%)
\end{array}\Biggr.
\\
&&\tilde N_3 \rightarrow \tilde N_1 Z\qquad (\sim 97\%)
\\
&&
\tilde C_1 \rightarrow \, \tilde t_1 b\qquad (\sim 95\%)
\eeq
for the Higgsino-like neutralinos and charginos. The wino-like neutralino 
and chargino $\tilde N_4$ and $\tilde C_2$ are so heavy in this class of 
models that they are unlikely to be directly produced in great numbers at 
any foreseeable colliders, but may appear (with small branching fractions) 
in decays of left-handed squarks. This scarcity of winos is 
different from the expectation in many mSUGRA models, for example. The 
left-handed squarks of the first two families decay predominantly through 
the gluino, while the right-handed squarks decay mostly directly to the 
LSP:
\beq
&&\tilde q_L \rightarrow 
\Biggl \{ \begin{array}{ll} 
q \tilde g \quad &(\sim 78\%)
\\
q' \tilde C_2\quad &(\sim 11\%)
\end{array}\Biggr.
\\
&&\tilde q_R \rightarrow\, q \tilde N_1 \qquad (\sim 90\%)
\eeq
However, the latter large branching fraction is due here to the very small 
phase space allowed for decays to the gluino, so while it is a real 
possibility, it cannot be considered a robust prediction for this class of 
models in general. 

The sleptons decay mostly directly to the bino-like LSP $\tilde N_1$.
Unfortunately, they almost completely decouple from the other superpartner
decay chains, so they must be produced directly to be observed. This
is clearly a challenge, given their large masses.

Because of the large masses of the entire superpartner spectrum, it will 
be difficult to probe the scenario outlined here at the Tevatron. The 
process
\beq
p \overline p \rightarrow \tilde t_1 \tilde t_1^* \rightarrow 
c \overline c \tilde N_1 \tilde N_1 \rightarrow c\,\overline c + \missingET
\eeq
is one of the searches being actively pursued by both the D$\emptyset$ 
\cite{Dzerostops} and CDF \cite{CDFstops} collaborations, using heavy 
flavor tags. However, the sensitivity appears to be well short 
\cite{Demina:1999ty} of that needed to probe most of the region favored by 
neutralino annihilations to top quarks as the solution for dark matter. 
The same process without the heavy flavor tag requirement, $p\overline p 
\rightarrow$ (acoplanar $jj$) $+ \missingET$ \cite{Dzeroacoplanarjets}, 
may also be interesting. However, in the present case the jets will not 
have a particularly high transverse momentum compared to the typical 
situation in mSUGRA benchmark models. The clean trilepton and $\missingET$ 
signal from wino-like $\tilde C_1 \tilde N_2$ 
production that is found in mSUGRA is 
unavailable here. The other superpartners generally are too heavy to be 
produced in sufficient numbers at the Tevatron with the anticipated 
integrated luminosity. The Higgs sector consists of a Standard-Model like 
Higgs boson $h^0$
and a heavy isotriplet of Higgs bosons ($H^0, A^0, H^\pm$), 
so enhanced signals are not 
expected there.

At the Large Hadron Collider, squark and gluino production will dominate 
as usual. The latter leads to the signal of two tops and two charm jets, 
but with 50\% probability for like-sign tops, because of the Majorana 
nature of the gluino decays:
\beq
pp \rightarrow \tilde g \tilde g \rightarrow 
\left \{ \begin{array}{ll}
t \,\overline t\, \tilde t_1\, \tilde t_1^* \rightarrow 
t \,\overline t\, c\, \overline c + \missingET\qquad (50\%)
\\
t \,t\, \tilde t_1^*\, \tilde t_1^*\rightarrow 
t \,t\, \overline c \,\overline c + \missingET\qquad(25\%)
\\
\overline t \,\overline t\, \tilde t_1\, \tilde t_1
\rightarrow 
\overline t\, \overline t\, c \, c + \missingET\qquad(25\%)
\end{array}
\right.
\label{eq:likesigntops}
\eeq
This LHC signal has been studied in \cite{Kraml:2005kb} using 
the like-charge
lepton signal arising from the
leptonic decay modes for both top quarks, with the result 
that both discovery and mass measurements are possible up to 
a gluino mass 
of about 900 GeV. The assumptions in the 
benchmark models 
used in that paper included a neutralino and stop that 
were both relatively lighter than in the scenario discussed in the present 
paper, but the results seem likely to be qualitatively the same.

Since many of the squarks produced at the LHC will decay through gluinos 
and then into top squarks and top quarks by eqs.~(\ref{eq:gluinodecay}), 
one expects also the same signal as in eq.~(\ref{eq:likesigntops}) with 
two extra (usually light-flavor) jets, some of which may however be 
relatively soft. Other squark-squark and gluino-squark events will yield 
the typical jets plus leptons plus $\missingET$ signatures.

For models like the one in Figure \ref{fig:spectrum}, sleptons will not 
appear with significant multiplicity in the decays of the squarks and 
sleptons that are produced directly at the highest rates at the LHC. For 
these masses, the direct production of sleptons would also be very 
difficult to observe in either Drell-Yan production \cite{LHCsleptonsDY} 
or vector boson fusion \cite{LHCsleptonsVBF}.

If the decay mode $\tilde t_1 \rightarrow W b \tilde N_1$ is open, as can 
occur for models with lower $|A_0|$ and/or higher $\Omega_{\rm DM} h^2$, 
then a different like-charge lepton signal results from $pp \rightarrow 
\tilde g \tilde g \rightarrow t \,t \overline b\, \overline b \,\ell^- 
\ell^- + \missingET$ and $\overline t \,\overline t\, b\, b\, \ell^+ 
\ell^+ + \missingET$ at the LHC. The resulting events with four potential 
$b$ tags, like-charge dileptons, and large missing energy should provide 
for a striking signal.

Finally, it is important to note that the scenario for neutralino LSP 
annihilation to a top quark does not present a particularly promising 
situation for a linear collider with $\sqrt{s} = 500$ GeV. In the model 
shown in Figure \ref{fig:spectrum}, and in a great deal of the allowed 
parameter space in Figure \ref{fig:LSPstop}, the only supersymmetric 
particle that can be produced at such a machine is 
$\tilde N_1$, which does not lead to a visible signal (except possibly 
through initial state radiation $e^+ e^- \rightarrow \tilde N_1 \tilde N_1 
\gamma$ \cite{ISRLSPs}). With that collision energy, only an $h^0$ with 
properties nearly indistinguishable from a Standard Model Higgs boson will 
be in direct evidence. Fortunately, if this is the course that Nature has 
chosen, the LHC should be able to identify the problem in advance, and 
allow for informed decisions regarding required beam energy. However, the 
difficulty in seeing sleptons at the LHC as noted above will present, as 
it does in many models, a challenge for assessing the capabilities of a 
linear collider.

\section{Outlook}\label{sec:outlook}
\setcounter{equation}{0}
\setcounter{footnote}{1}

In this paper, I have argued that there is a particularly appealing region 
of parameter space in which the right amount of dark matter can be 
obtained naturally, in the sense that $\Omega_{\rm DM} h^2$ varies rather 
slowly with changes in the input parameters. The key feature is 
suppression of $\Omega_{\rm DM} h^2$ due primarily to neutralino pair 
annihilation to top quarks through top squark exchange, which can occur 
with moderate top-squark mixing in models that have a relatively light 
gluino compared to the predictions of models with universal gaugino mass 
boundary conditions as in mSUGRA. The resulting superpartner spectrum can 
therefore be described as compressed, and leads to rather distinctive 
predictions. The fact that the $\tilde N_1$ mass
has to exceed the top quark 
mass in this scenario for dark matter makes the discovery of supersymmetry 
impossible\footnote{This is hardly a bold prediction now, but it does have 
the virtue of being inevitable in this scenario.} at the past CERN LEP 
$e^+ e^-$ collider, very difficult for the ongoing Tevatron, but quite 
promising for the LHC. Since the $\tilde t_1$ squark has to be still 
heavier by tens of GeV, there is considerable doubt that a linear collider 
would be able to see it, or any other superpartners, unless the 
center-of-mass energy is higher than $\sqrt{s} = 500$ GeV.

In this paper, I have not attempted any detailed study of LHC potential 
for discovery or precision measurements, or of the possibilities for 
direct or indirect detection of the dark matter. These issues will 
hopefully be addressed in future work.

{\bf Acknowledgments:} I am grateful to James Wells for useful
conversations. This work was supported by the National Science 
Foundation under Grant No.~PHY-0456635.

\end{document}